\documentclass{article}

\usepackage[nonatbib ,preprint]{neurips_2023}

\usepackage[utf8]{inputenc} 
\usepackage[T1]{fontenc}    
\usepackage{hyperref}       
\usepackage{url}            
\usepackage{booktabs, makecell}       
\usepackage{amsfonts}       
\usepackage{nicefrac}       
\usepackage{microtype}      
\usepackage[table, svgnames]{xcolor}         
\usepackage{tabularx}
\usepackage{multirow}
\usepackage{subfigure}
\usepackage{caption}
\usepackage[pdftex]{graphicx}
\usepackage{biblatex}
\usepackage{adjustbox}
\usepackage{amsmath} 
\usepackage{amssymb}
\pdfoutput=1
\title{MAGMA: Music Aligned Generative Motion Autodecoder }

\author{%
  Sohan Anisetty \\
  Department of Computer Science\\
  Georgia Institute of Technology\\
  \texttt{sanisetty3@gatech.edu} \\
  \And
  Amit Raj \\
  Department of Computer Science \\
  Georgia Institute of Technology \\
  \texttt{amit.raj@gatech.edu} \\
  \AND
  James Hays \\
  Department of Computer Science \\
  Georgia Institute of Technology \\
  \texttt{hays@gatech.edu}
}

\begin{document}

\maketitle

\begin{abstract}
Mapping music to dance is a challenging problem that requires spatial and temporal coherence along with a continual synchronization with the music's progression. Taking inspiration from large language models, we introduce a 2-step approach for generating dance using a Vector Quantized-Variational Autoencoder (VQ-VAE) to distill motion into primitives and train a Transformer decoder to learn the correct sequencing of these primitives. We also evaluate the importance of music representations by comparing naive music feature extraction using Librosa to deep audio representations generated by state-of-the-art audio compression algorithms. Additionally, we train variations of the motion generator using relative and absolute positional encodings to determine the effect on generated motion quality when generating arbitrarily long sequence lengths. Our proposed approach achieve state-of-the-art results in music-to-motion generation benchmarks and enables the real-time generation of considerably longer motion sequences, the ability to chain multiple motion sequences seamlessly, and easy customization of motion sequences to meet style requirements.


\end{abstract}

\section{Introduction}
\label{intro}

Music has always played an important role in motivating and propelling movement, from traditional folk dances to recent contemporary performances. In recent years, there has been an increasing interest in the intersection of music and technology, particularly in the area of automated dance generation. The complexity of dance movements and the constant changes in musical elements make it difficult to create an automated system for music-conditioned dance generation. While there has been extensive research on text-to-motion generation, these methods are not always suitable for music conditioned motion generation. The complex nature of dance requires not only spatial and temporal coherence but also a continual synchronization with the music as it progresses contrasting with the one-shot nature of text-based approaches. Various complex methods have been proposed to enforce this synchronization, ranging from mapping music and dance to a unified latent space and learning rhythm signatures\cite{choreomaster}, using reinforcement learning to align dance with music beats\cite{bailando}, and using multiple encoders, discriminators, and decoders\cite{dancing2music} to generate dance resembling the ground truth. Further, many popular techniques use a seed motion\cite{aichoreo,youneverstop, dancerevolution} and also require access to information about music features at future timesteps which might not be available during in-the-wild inference and hinders real-time motion generation. 
However, recent advancements in generative AI, especially in the image generation space\cite{vqvae2,vqgan,stablediff,hierarchicaldiffusionclip} have been brought over to the motion generation field\cite{mdm,t2mgpt,edge} proving that simpler architectures with well designed optimization goals are sufficient to achieve a high level of synchronization between the conditioning input and generated motion. We take inspiration from large language models, especially multi-modal architectures\cite{ofa,flamingo,mPlug,visualgpt,} which are task and modality agonostic and beat task curated models on image captioning, visual grounding and visual question answering tasks. We follow a similar ideology and train our motion generation pipeline on a combined dataset of text-to-motion and music-to-motion datasets to encode a richer motion representation and offer stronger generalization to the type of conditioning inputs used. Our results corroborate the results observed in the vision-language domain\cite{ofa,beit3}, training on such a combined dataset gives improved performance in both text-to-motion and music-to-motion tasks compared to models trained specifically only on music or text.

Our framework incorporates a Vector Quantized-Variational AutoEncoder (VQ-VAE)\cite{vqvae} to condense general human motion into "motion primitives" analogous to text tokenization, and an autoregressive Generative Pretrained Transformer (GPT)\cite{gpt} with causal attention to sequentially construct motion sequences conditioned on the input music and optional textual style. These motion primitives correspond to VQ-VAE codebook indices while music is represented as Librosa\cite{librosa} or Encodec\cite{encodec} music features. Prior work on long-form generation\cite{aichoreo,youneverstop} posit that without information about future frames the model collapses to generate small-magnitude motion and eventually freezes up. However, we observe that our model is able to generate motion sequences greater than 4 times the maximum length seen during training using only previous timestep information without collapsing and not freezing up. Through our proposed approach, we achieve results that are comparable to the current state-of-the-art methods in general motion reconstruction\cite{t2mgpt} and obtain state-of-the-art results on music-to-motion generation tasks on the AIST++\cite{aichoreo} dataset. 

In summary, our contributions are the following:

1. We introduce a 2 step approach for generating dance using a VQ-VAE to distill motion into primitives and train a modified Transformer decoder to learn the correct sequencing of these primitives.

2. We evaluate the importance of music representations by comparing naive music feature extraction using Librosa\cite{librosa} to deep audio representations generated by state-of-the-art audio compression algorithms\cite{encodec,soundstream,jukebox}.


3. We introduce an additional textual conditioning input inspired by StyleGAN\cite{stylegan2} which can open up possibilities of generating dance conditioned primarily on music but influenced also by textual prompts such as mismatched genre or pace.

\section{Related Work}
\label{related_work}

\subsection{Vector Quantization} The Vector Quantized Variational Autoencoder(VQ-VAE), was proposed in \cite{vqvae} as an extension to VAE \cite{vae} by learning a discrete latent space instead of a continuous normal distribution. The VQ-VAE encoder uses a predetermined codebook to compress the input data into discrete latent codes, and the decoder then reconstructs the original data from these codes. This discrete representation of the latent space encourages the model to construct a structured and interpretable representation that effectively captures the salient characteristics of the data. VQ-VAE's have shown promising results in generative tasks across various domains, such as image synthesis \cite{hierarchical, vqgan, vqvae2}, text-to-image generation \cite{zeroshotvqvae}, and audio compression and generation\cite{jukebox, audiolm, musiclm, soundstream, encodec} Building on this success, VQ-VAE's have been used to model motion\cite{t2mgpt, bailando} successfully by using a 2 stage approach; encoding motion data into a discrete space and then learning a probabilistic model to generate motion indices.  

\subsection{Human motion generation}

Generating human motion sequences that are both spatially consistent and temporally coherent is a challenging task, and can be guided by different types of conditioning inputs such as action class\cite{actor,humanml3d}, audio\cite{aichoreo,bailando,edge,choreomaster,lietal,dancerevolution,dancing2music,music2dance}, and natural language\cite{t2mgpt,teach,motionclip,motiondiffuse,mdm,temos,tm2t}. While some works focus on unconditional motion synthesis\cite{modi} or motion editing\cite{singleshot, harvey2021recurrent, robustinbetween}, most current methods suggest dedicated approaches to map each conditioning domain into motion. In contrast, our approach aims to learn a unified architecture that can generate plausible motion sequences irrespective of the conditioning input domain.

\paragraph{Text conditioned motion synthesis}

ACTOR\cite{actor} proposes a transformer-based VAE that generates the entire motion sequence in one shot conditioned on action embeddings sampled from a learned normal distribution. TEMOS\cite{temos} and T2M\cite{humanml3d} extend ACTOR by using a Variational Autoencoder (VAE)\cite{vae} to map text descriptions to a normal distribution in latent space, which are then used instead of the action embeddings. MotionCLIP\cite{motionclip} learns a joint text-image-motion latent space by training an autoencoder to align motion sequences to text and its rendered image. By taking advantage of CLIP's rich semantic representation it is able to generate out of domain motions and opens up latent space editing capabilities. MDM\cite{mdm} and MotionDiffuse\cite{motiondiffuse} use diffusion-based models for text-to-motion generation using classifier free guidance\cite{classifierfree} and showcase motion editing and infilling capabilities. T2M-GPT\cite{t2mgpt}, which is most similar to our work, employs a two-stage approach to decompose motion into motion primitives represented by codebook indices and uses a GPT\cite{gpt} motion generator conditioned on CLIP\cite{clip} text embeddings to autoregressively generate plausible combinations of index sequences. However, they use CLIP\cite{clip} text encodings as the first token to prompt autoregressive motion generation while music embeddings need to continuously influence motion generation. 


\paragraph{Music conditioned dance generation}

Early approaches explore matching existing 3D motion to music using motion retrieval\cite{msicconvsynt,learn2dance,musicsim}, but the resulting choreographies lacked the complexity of human dances and could not generate new motions beyond the available database. A promising approach is prediction-based methods, which treat dance generation as a motion prediction problem. Various network architectures have been proposed, including CNN\cite{dancing2music,motionet,charmotsyn,learning2dance_gcn,bailando,music2dance}, RNNs/LSTMS\cite{groovenet,tempguidemusic,dancemelody,rhythmisadancer}, GANs\cite{dancing2music,deepdance}, reinforcement learning\cite{bailando}, motion graphs\cite{choreomaster}, diffusion models\cite{edge} and Transformers\cite{genrecond,aichoreo,youneverstop,attenspatiotemporal,dancerevolution}. However, many of these approaches need specialised pre-processing to work with in-the-wild music\cite{bailando}, require a seed motion and the entire music sequence in hand\cite{aichoreo}, or have complex architectures\cite{choreomaster, dancing2music,danceformer} necessitating the need for a simpler alternative, that can generalize well to in-the-wild scenarios. EDGE\cite{edge} modifies diffusion based text-to-motion generation\cite{mdm} by cross-attending to jukebox\cite{jukebox} music embeddings. They generate 5 second motion sequences and accomplish long-form generation by and stitching them together and enforcing consistency for overlapping 2.5 second slices. We propose a autoregressive transformer-based method for generating novel motion sequences that can be conditioned on both arbitrary length music and textual styles, and is capable of generalizing well to in-the-wild scenarios.

\section{Preliminaries}
\label{section:prelims}

\paragraph{Pose Representation:} We use the HumanML3D\cite{humanml3d} dataset which has 22 SMPL\cite{smpl} joints represented by \( x \in \mathbb{R}^{d_h}\), where \(d_h\) is the joint representation dimension corresponding to joint local and global rotations/positions, velocities, and binary foot contact features at 20 frames per second (FPS). AIST++\cite{aichoreo} music-to-dance dataset uses 24 SMPL\cite{smpl} joints with 9DOF rotation representation for every joint and root translation at 60FPS. We downsample and apply the HumanML3D pre-processing steps on AIST++ in order to build a combined dataset of text-dance-motion at 20FPS with 22 joints. 

\paragraph{Conditioning Representation:} AIST++\cite{aichoreo} music features \(c^a\) are obtained using Librosa\cite{librosa} and consist of of MFCC, chroma, envelope, one-hot peaks, and one-hot beats features. We also experiment with Encodec\cite{encodec} embeddings \(c^e\) as the conditioning signal for motion generation. We resample raw audio at 24KHz to 6.4KHz resulting in a compressed representation at 20Hz matching the motion framerate. We use CLIP\cite{clip} to extract text embeddings, which has been widely used in research to guide image and motion generation conditioned on text\cite{motionclip, styleclip,hierarchicaldiffusionclip,t2mgpt,mdm}, for style modulation.

\begin{figure}
  \centering
  \includegraphics[width = 0.6\linewidth ]{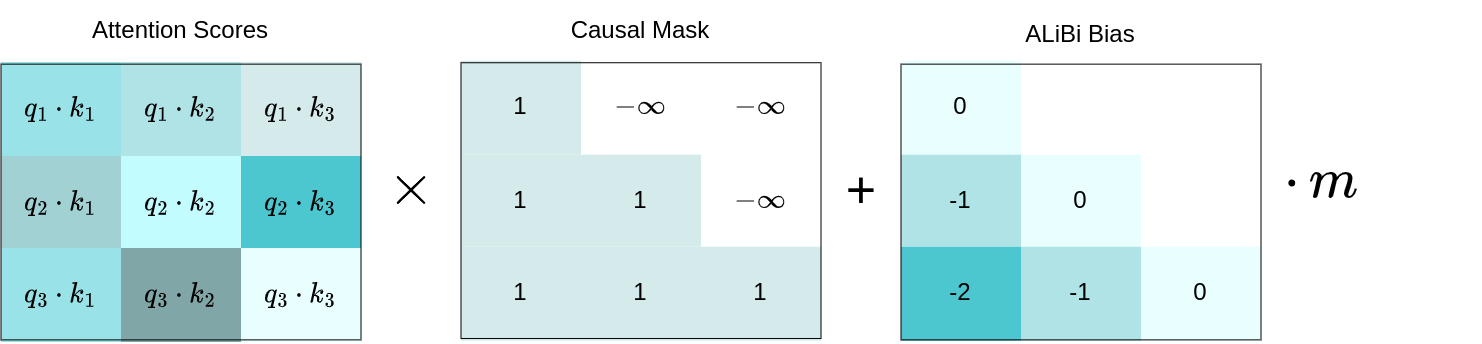}
  \caption{We compute the attention scores using a causal mask and a constant bias\cite{ALiBi} with a predefined slope hyperparameter m.}
  \label{fig:attention}
\end{figure}

\paragraph{Causal attention and relative positional encodings:} 
Current research on text-to-motion or music-to-dance generation use absolute positional encodings to generate motion sequences that are often short in length or restricted to the maximum length seen during training. Absolute positional encodings show limited generalization to longer sequence lengths as demonstrated in \cite{ALiBi, roformer}. Therefore, we compare the performance of absolute embeddings with relative positional encodings\cite{ALiBi} to allow for longer, one-shot generation without the need to generate multiple smaller sequences and concatenate them\cite{edge}. The causal attention with AliBi\cite{ALiBi} is formulated in Equation \ref{eq:attention} and depicted in Figure \ref{fig:attention}:

\begin{equation}
\label{eq:attention}
    A_c(Q,K,V)= Softmax \left( \frac{QK^T  \times m_c + m \cdot bias}{\sqrt{d_k}} \right) \cdot V
\end{equation}

Causal Attention \(A_c\) is computed as a Softmax function applied to the product of Query \(Q \in \mathbb{R}^{T \times d_k}\) and Key \(K \in \mathbb{R}^{T \times d_k}\) matrices, which are multiplied by a causal mask \(m_c\) and summed with a bias term derived from the ALiBi\cite{ALiBi}. The causal mask is applied to restrict the model's attention only to previous states. Specifically, the values of the causal mask are set to negative infinity if the row index is greater than the column index, and 1 otherwise. Here, \(d_k\) represents the embedding dimension of the transformer, and the slope \(m\) is a predetermined scalar specific to each attention head\cite{ALiBi}.

\section{Method}
\label{method}

Our objective is to generate motion that aligns spatially and temporally with the input music, while also being applicable to genres and styles beyond its training data. Drawing inspiration from visual-language architectures\cite{mPlug,flamingo,univl,coarsetofine}, we devise a motion generation pipeline cross-conditioned on music (Figure \ref{fig:full_pipeline}). We pretrain on a mixed dataset with the text-to-motion HumanML3D\cite{humanml3d} dataset to augment the sparse training data in the AIST++\cite{aichoreo} music-to-dance dataset. In our approach, we employ a VQ-VAE to discretize motion in the combined dataset into discrete motion primitives using a learned codebook and utilize the transformer architecture to autoregressively generate codebook indices conditioned on music embeddings. 


\subsection{VQ-VAE}
\label{section:vqvae_method}

\begin{figure}
  \centering
  \includegraphics[width = \linewidth]{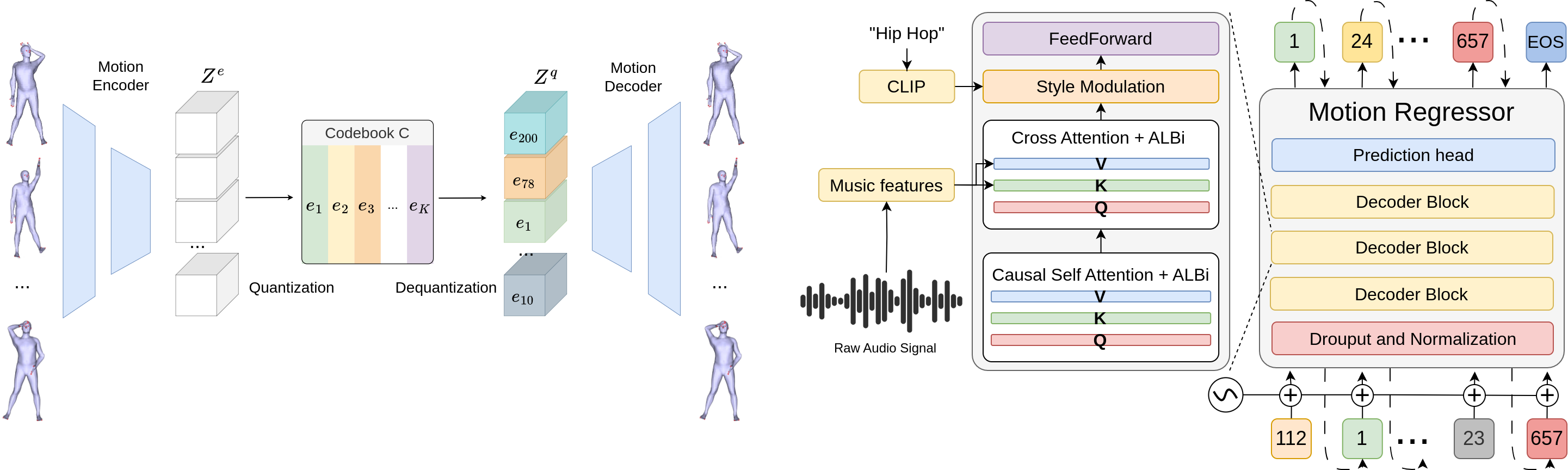}
  \caption{\textbf{Our 2 stage motion generation pipeline:} The VQVAE module shown on the left distills human motion into primitives represented by a codebook \(C\). Analogous to tokenization in NLP, we use the codebook indices as input tokens to the motion regression module shown on the right. We train an autoregressive decoder cross-conditioned on music features and modulated on style (text embeddings) that is capable of extrapolating to long sequence lengths using relative positional embeddings (ALIBI)\cite{ALiBi}.}
  \label{fig:full_pipeline}
\end{figure}

The VQ-VAE\cite{vqvae} framework has been successfully used to encode complex high dimensional data such as audio\cite{soundstream, encodec, jukebox} and images\cite{vqvae2, vqgan} into discrete representations. Building on these works, we aim to learn a codebook \(C\) consisting of embeddings \(\{e_k \in \mathbb{R}^{d_c}\}^K_{k=1} \), where K is the number of codebooks with dimension \(d_c\) such that a motion sequence with \(T\) frames, \(X = [x_1,x_2,....,x_T] \) with \(x_t \in \mathbb{R}^d\), can be reconstructed back after passing through the autoencoder architecture and discretized by the codebook as shown in Figure \ref{fig:full_pipeline}. Passing the motion sequence \(X\) through the Encoder \(E\) results in latent features \(Z^e = E(X)\), with \(Z^e = [z^e_1,z^e_2,....,z^e_T] \) and \(z^e \in \mathbb{R}^{d_c}\). 
\paragraph{training objective:} For i-th latent feature \(z^e_i\), the quantization through the codebook is to find the most similar element in \(C\), which can be written as:

\begin{equation}
z^q_i = \underset{c_k \in C}{argmin} ||z^e_i - e_k ||_2
\end{equation}

A decoder \(D(e)\) then decodes the embedding vectors back into the input space. The original formulation of the optimization goal\cite{vqvae} is:

\begin{equation}
\mathcal{L}_{vq} = \underbrace{L_{huber}(X , D(Z^q))}_{reconstruction}  + \underbrace{|| sg[Z^e] - Z^q||_2}_{codebook} + \underbrace{\beta||Z^e - sg[Z^q]||_2}_{commit}
\end{equation}

Where \(sg\) stands for the stop-gradient operator that has zero partial derivatives during back-propogation. The codebook entries are optimised solely by the codebook loss while the commit loss prevents the encoder output from growing arbitrarily by constraining the encoder to the codebook embedding space. \(L_{huber}\) corresponds to the huber loss between the input and reconstructed motion sequences. 

We found that training using this formulation results in codebook collapse, where a significant number of encodings getting mapped to a smaller subset of codebook vectors. We use the training tricks outlined in follow up work\cite{vqvae2,vqgan,jukebox,soundstream} to increase codebook usage. \textbf{Random restart}\cite{jukebox,soundstream} replaces stale codebook vectors with a random embeddings from \(Z^e\). \textbf{Exponential moving average (EMA) updates}\cite{vqvae} speeds up training by updating each codebook vector using an average of its \(n\) nearest embeddings in the input. \textbf{Kmeans initialization}\cite{soundstream} initializes the codebook by the kmeans centroids of the first batch. We use the transformer\cite{transformers} encoder architecture for the VQ-VAE encoder and a causal decoder\cite{gpt} for the VQ-VAE decoder with a modified attention mechanism formulated in Equation \ref{eq:attention}. 


\subsection{MotionSeq Decoder}
\label{section:gpt_method}
\paragraph{Optimization goal.} We utilize the VQ-VAE model to encode the input motion sequence \(X = [x_1,x_2,....,x_T] \) into a sequence of indices \(S = [s_1,s_2,....,s_T] \) that is appended with a special \textit{<EOS>} token to indicate the end of a sequence. Given input \(S\) and condition \(c = [c_1, c_2, ..., c_{T}]\), we train the decoder by minimizing the negative log likelihood of the predicted data distribution 
\begin{equation}
    \mathcal{L}_{gpt} = -\sum_{i=1}^{|S|} log[P_{\theta}(S_i|S_{<i} , c)]
\end{equation}

where \(\theta\) refers to the model parameters. In contrast to text generation models, which use a \textit{<BOS>} (Beginning Of Sequence) token as the first token, we use the first index \(s_1\) to facilitate easy concatenation of multiple generated sequences. Prior music based dance generation research\cite{aichoreo,bailando,dancing2music,music2dance} use acoustic features (Section \ref{section:prelims}) while \cite{edge,dancerevolution} use features from music based networks like Jukebox\cite{jukebox} citing greater semantic representative capabilities. We perform ablation studies on these audio representations in Section \ref{section:discussion}. We have 4 major components in each decoder layer of the autoregressive network as shown in Figure \ref{fig:full_pipeline}, a causal self-attention layer, a cross-attention layer, a style modulation layer, and a feedforward dense layer. 

\paragraph{Cross Attention}: In our implementation, we use cross-attention in addition to causal self attention\ref{eq:attention}, where the Key and Value embeddings are computed using external conditioning inputs, in our case music features. We use a single feed-forward layer to project these embeddings to the transformer dimension \(d_k\). 

\paragraph{Style Modulation.}The style modulation layer takes inspiration from adaptive instance normalization in StyleGAN\cite{stylegan,stylegan2} to mix textually defined styles with transformer attention outputs. Formally,
\begin{equation}
    \mathrm{Modulation}(A_c,\xi) = \xi \cdot \frac{A_c}{||A_c||_2},\, A_c \in \mathbb{R}^{T \times d_k}, \xi \in \mathbb{R}^{d_s}
\end{equation}
where \(A_c\) refers to the output of the previous layer and \(\xi\) is the CLIP\cite{clip} embedding of textual style. We use a single feed-forward layer to project the style embedding dimension \(d_s\) to transformer dimension \(d_k\). We do not include this layer by default and perform an ablation study on the inclusion of style modulation in Section \ref{section:discussion}. 

During training, given an input sequence \([s_1,s_2,....,s_{T}]\), corresponding condition \(c = [c_1, c_2, ..., c_{T}]\), and a style embedding \(\xi\), we aim to predict the target sequence \([s_2,s_3,....,s_T, EOS]\). During inference, we begin with random codebook indices and music embeddings of the same sequence length as input tokens. We then generate indices in an auto-regressive manner, incrementally increasing the sequence length of the music condition until the target length is reached. 


\section{Experiments}
We compare the performance of our two step pipeline with recent research\cite{t2mgpt,music2dance,dancerevolution,aichoreo,bailando,edge} (Section \ref{section:results_quant}) on standard datasets (Section \ref{section:datasets}) and widely used metrics (Section \ref{section:eval_metrics}).

\subsection{Datasets}
\label{section:datasets}

\textbf{AIST++.} The AIST++\cite{aichoreo} dataset contains 1408 dance sequences performed by 30 dancers paired with music from 10 genres with basic and advanced choreographies at 60FPS. The dataset has motion sequence lengths ranging from 7 seconds to 50 seconds with an average length of 13 seconds.

\textbf{HumanML3D.} The HumanML3D\cite{humanml3d} dataset contains 14,616 human motion at 20FPS and 44,970 text descriptions. Pre-processing used to generate the dataset is also applied to AIST++ motions to form a combined dataset with 32048 motion sequences spanning 86 hours. 

\subsection{Implemetation details}
We use a codebook of size K = 1024, with embedding dimension \(d_c\) = 768 in the VQ-VAE. Both the VQ-VAE encoder-decoder and motion regressor have a depth of 12 layers, with 8 attention heads and a dimension of \(d_k = 768\). The HumanML3D dataset has a pose representation of \(d_h\) = 768, Librosa music features have dimension \(c^a\) = 35, and Encodec\cite{encodec} features \(c^e\) has a dimension of 128. The VQVAE is trained for 300k steps while all variations of the motion regressor are trained for 200k steps. We use the AdamW\cite{adamw} optimizer, batch size of 128 for VQ-VAE and 64 for motion regressor, learning rate of 2e-4, and a cosine learning rate schedular with a warmup of 5000 steps. We set the commit loss weight \(\beta\) to 0.2, and the head-specific slope \(m\)\cite{ALiBi} is defined as the geometric sequence that commences at \(2^{\frac{-8}{n}}\) and utilizes the same value for its ratio for \(n\) heads. We train all variations of the models on 2 NVIDIA A40 GPUs using HuggingFace\cite{huggingface} accelerate and fp16 mixed precision training.


\subsection{Evaluation Metrics}
\label{section:eval_metrics}

We follow the evaluation methodology specified in \cite{humanml3d} and \cite{aichoreo} which has been widely used in follow up research. We use pre-trained networks\cite{humanml3d} to extract motion and text representations to calculate text-motion metrics and expert designed\cite{aichoreo} geometric and kinetic feature extractors for music-dance metrics. All evaluation metrics computed in Table \ref{table:vqvae_results} and Table \ref{table:gpt_results} are repeated 20 times.

\textbf{Frechet Inception Distance (FID):} We calculate the distribution distance between the generated and real motion using FID\cite{fid} on the extracted motion features. \({\mathrm{\textit{FID}}}_g\) evaluates geometric relations between joints across the generated sequence and \(\mathrm{\textit{FID}}_k\) evaluates kinetic aspects of the motion.

\textbf{Diversity:} We randomly select motion pairs from the test set and calculate their motion features to compute the average Euclidean distance. Our observation is that jittery motion tends to result in diversity scores higher than ground truth. Therefore, we consider the models that match the real motion diversity more closely to have better performance.

\textbf{Beat Alignment:}
We evaluate the motion-music correlation in terms of the similarity between the kinematic beats and music beats. The music beats are extracted using Librosa\cite{librosa} and the kinematic beats are computed as the local minima of the motion joint velocity.

\textbf{R-Precision:} Given one motion sequence and 32 text descriptions (1 ground-truth and 31 randomly selected mismatched descriptions), we rank the Euclidean distances between the motion and text embeddings. Top-1, Top-2, and Top-3 accuracy of motion-to-text retrieval are reported. 

\textbf{Multimodal Distance:} The average Euclidean distances between each text feature and the generated motion feature from this text.

\subsection{Results}
\label{section:results_quant}
We compare our method to \cite{t2mgpt, dancerevolution,music2dance,aichoreo,bailando,edge} in motion reconstruction and dance generation tasks on HumanML3D\cite{humanml3d} and AIST++\cite{aichoreo} datasets respectively.

\begin{figure}
  \centering
  \includegraphics[width = 0.8\linewidth]{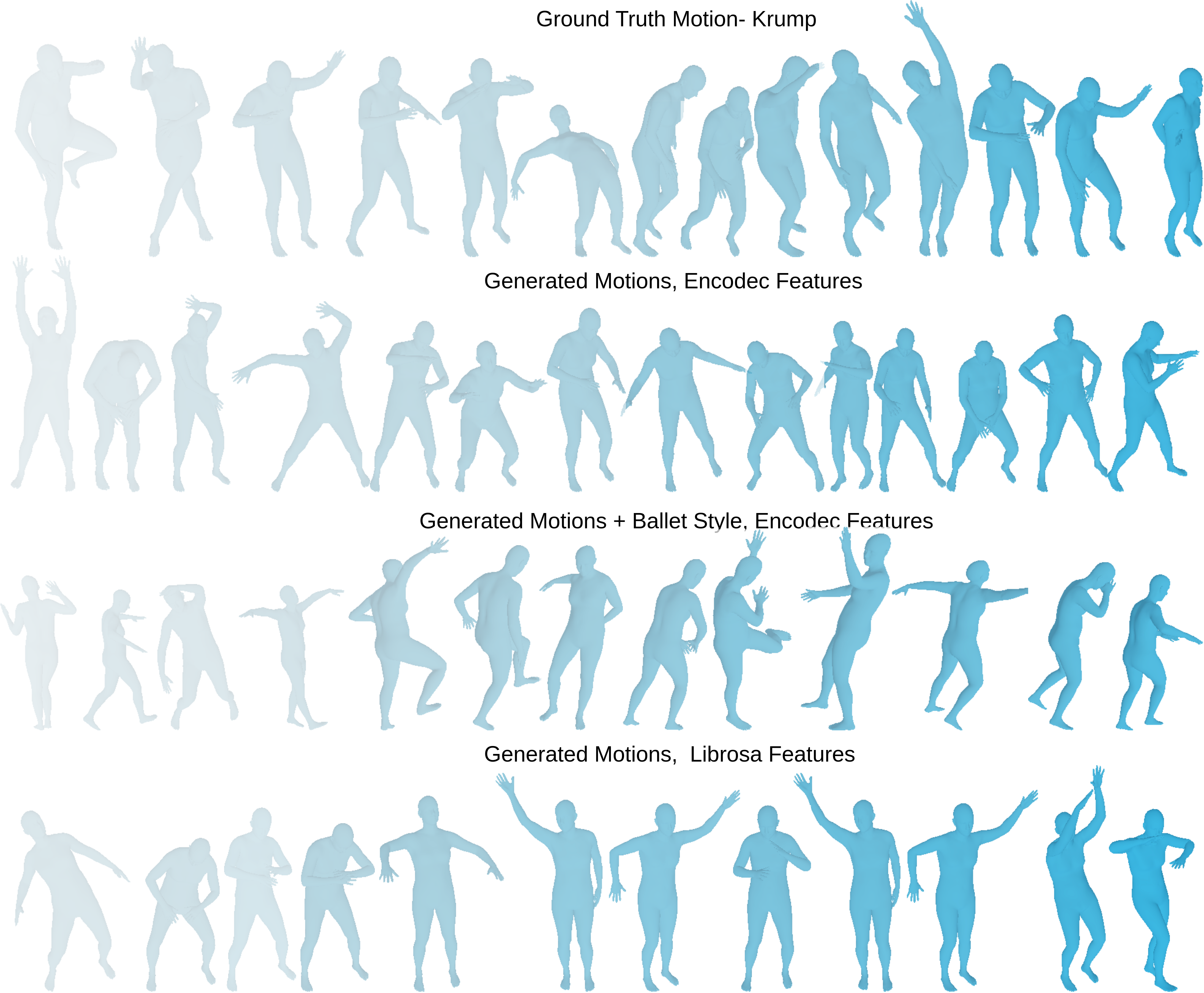}
  \caption{Our model produces semantically similar results (row 2) on the "Krump" music-dance genre, known for its pronounced upper body motions (top row). When conditioned on the "Ballet" genre, our model seamlessly incorporates long strides reminiscent of ballet (Row 3), while maintaining the essence of the original style. Librosa\cite{librosa} music features lead to consistent motions in repeated music segments, whereas using Encodec\cite{encodec} features result in a more diverse range of motions.}
  \label{fig:comparision}
\end{figure}

\subsubsection{Quantitative results}

We showcase results of our VQ-VAE model on the HumanML3D\cite{humanml3d} dataset and AIST++\cite{aichoreo} dataset in Table \ref{table:vqvae_results} and Table \ref{table:gpt_results} respectively. We use 3 variations of our model, \textit{Only T2M} denotes training only on the HumanML3D dataset. \textit{Only T2M + Only AIST} denotes pre-training only on the HumanML3D dataset and finetuned only on the AIST++ dataset for an equal number of epochs. \textit{T2M + AIST} denotes pre-training on the combined dataset of the HumanML3D and AIST++ dataset and fine-tuning for a few epochs on the AIST++ dataset. We observe that training on the combined dataset achieves comparable performance on the HumanML3D\cite{humanml3d} dataset to T2M-GPT\cite{t2mgpt} that was trained solely on it, while offering a significantly closer diversity score to the real motion. We test the performance of our VQ-VAE model variations on the AIST++ dataset as shown in Table \ref{table:gpt_results}. Surprisingly the model trained solely on the text-to-motion dataset\cite{humanml3d} was still able to generalise reasonable well to dance. The \textit{Only T2M + Only AIST} model achieves the lowest FID but suffers in diversity and beat align scores, indicating that the codebook memorised the dance sequences in the train split of the dataset. \textit{T2M + AIST} achieves the best diversity scores (\(Dist_k\) and \(Dist_g\)) and better correlation to music. To evaluate motion generation we autoregressively generate motion with \(T = 400\) frames (20 secs) using music specified in the test set (Table \ref{table:gpt_results}). For a fair evaluation we transform our motion representation back to the original AIST++\cite{aichoreo} one. We observe that our model was able to generate continuous dance without freezing artifacts, and minimal gliding, problems prevalent in previous work\cite{aichoreo,bailando}. We achieve the lowest FID scores and significantly closer diversity to the real motion while aligning with music effectively.  We were also able to generate longer sequences (upto 800 frames) in one shot before inconsistencies and excessive jitter was observed. Edge\cite{edge} observed that their \(\mathrm{FID}_k\) values were not consistant with their qualitative results and hence do not include them.

\subsubsection{Qualitative results}

Figure \ref{fig:comparision} presents qualitative results of our motion generation pipeline. Our model produces motion sequences where the dancer moves dynamically in the space with appropriate foot movements, distinguishing it from other methods that generate more stationary motion\cite{aichoreo,bailando,dancerevolution}. Incorporating style conditioning leads to notable modifications in the generated motion. For instance, when adding the "Ballet Jazz" style, the dancer intermittently performs ballet steps while following the original music clip. We observe that previous models\cite{aichoreo,bailando} experience motion freezing and noticeable foot gliding artifacts as the generation progresses. In our work, we have reduced these issues, resulting in smoother and more natural-looking motion. Video samples are in the supplementary materials.



\begin{table}
  \caption{VQ-VAE motion reconstruction results on HumanML3D\cite{humanml3d} test set. \(\uparrow/ \downarrow\) means bigger/smaller the better and \(\rightarrow\) means closer to real values the better. }
  \label{table:vqvae_results}
  \centering
    \begin{adjustbox}{width=1\textwidth}


  \begin{tabular}{lllllll}
    \toprule
    Model & \multicolumn{3}{c}{R-Precision}$\uparrow$  &FID $\downarrow$ &MM-Dist $\downarrow$   & Diversity $\rightarrow$                 \\
    \cmidrule(r){2-4}
         & Top-1 &Top-2 &Top-3 &  & &  \\
    \midrule
    \noalign{\vskip-\aboverulesep}
    \rowcolor{WhiteSmoke!70!Lavender} 
    Real motion & 0.511  & 0.703 & 0.797 & 0.002 & 2.974 & 9.503    \\
    T2M-GPT\cite{t2mgpt} & 0.501  & 0.692 & 0.785 & 0.07 & 3.072 & 9.593 \\
    Ours (Only T2M) & 0.504 & \textbf{0.699} & 0.7894 & 0.0689 & 3.011 & 9.858    \\
    Ours (Only T2M + Only AIST ) & 0.372  & 0.5721 & 0.6963 & 3.22 & 4.02 & 7.522    \\
    Ours (T2M + AIST) & \textbf{0.51}  & 0.694 & \textbf{0.7951} & \textbf{0.06} & \textbf{3.02} & \textbf{9.46}    \\
    
    \bottomrule
  \end{tabular}
    \end{adjustbox}

\end{table}

\begin{table}
  \caption{Motion generation results on the AIST++\cite{aichoreo} test set.}
  \label{table:gpt_results}
  \centering
  

  \begin{adjustbox}{width=1\textwidth}
  \begin{tabular}{llllll}
    \toprule
    Model & $\mathrm{FID}_k$ $\downarrow$ & $\mathrm{FID}_g$ $\downarrow$ & $\mathrm{Dist}_k$ $\rightarrow$ & $\mathrm{Dist}_g$ $\rightarrow$ & Beat Align $\uparrow$            \\
    \midrule
    \noalign{\vskip-\aboverulesep}
    \rowcolor{WhiteSmoke!70!Lavender} Real motion & --  & -- & 9.50 & 7.55 & 0.243 \\
    Our VQVAE (Only T2M) & 5.29 & 10.78& 8.2 & 7.0 & 0.2    \\
    Our VQVAE (Only T2M + AIST ) & \textbf{2.09}  & 7.25 & 8.91 & 7.41 & 0.274  \\
    Our VQVAE (T2M + AIST) & 2.42  & \textbf{6.61} & \textbf{9.06} & \textbf{7.44} & \textbf{0.286}    \\

    \toprule
    Dancenet \cite{music2dance} & 69.18  & 25.49 & 2.86 & 2.85 & 0.143 \\
    DanceRevolution \cite{dancerevolution} & 73.42  & 25.92 & 3.52 & 4.87 & 0.1950 \\
    FACT \cite{aichoreo} & 35.35  & 22.11 & 5.94 & 6.18 & 0.221 \\
    Bailando \cite{bailando} & 28.16  & \textbf{9.62} & 7.83 & 6.34 & 0.233\\
    \(\mathrm{EDGE}\) \cite{edge} & --  & 23.08 & \textbf{9.48} & {5.72} & 0.26 \\
    Ours (Encodec) & \textbf{10.34} & {11.16} & {10.06} & \textbf{7.78} & \textbf{0.26}\\

    \bottomrule
  \end{tabular}
\end{adjustbox}

\end{table}

\section{Discussion}
\label{section:discussion}
For ablation studies we train all variants of the model with a maximum sequence length of 300 frames (15s) and evaluate metrics on motion generated till 800 frames (40s). Since long duration motions are not available in the test set, we use ground truth motions from the train set for evaluation. We do not map the motions to the Aist++ representation.
  

\begin{table}
  \caption{Ablation studies on the AIST++\cite{aichoreo} train subset}
  \label{table:abl_studies}
  \centering
  
  \begin{tabular}{llllll}
    \toprule
    Model & $\mathrm{FID}_k$ $\downarrow$ & $\mathrm{FID}_g$ $\downarrow$ & $\mathrm{Diversity}_k$ $\rightarrow$ & $\mathrm{Diversity}_g$ $\rightarrow$ & Beat Align $\uparrow$            \\
    \midrule
    \noalign{\vskip-\aboverulesep}
    \rowcolor{WhiteSmoke!70!Lavender}
    Real motion & --  & -- & 9.50 & 7.55 & 0.243  \\
    Ours (Encodec) & {2.23} & 7.83 & 10.21 & 7.21 & 0.23\\
    Ours (Encodec, Abs pos) & 4.8 & 9.6 & 10.846 & 7.42 & 0.225\\
    Ours (Encodec + style) & 3.522 & 8.57 & 10.22 & 7.215 & 0.25\\
    Ours (Librosa features) & 3.32 & 8.37 & 9.64 & 7.19 & 0.25\\
    \bottomrule
  \end{tabular}
\end{table}


\textbf{Codebook.} We observe that using \textit{Random restart}, \textit{EMA updates}, and \textit{Kmeans initialization} provide a 20\% improvement in codebook usage compared to the naive implementation. However, a considerable number of codebook embeddings are still unused even when trained on the mixture of HumanML3D and AIST++ dataset. Thus, a smaller codebook might still be able to represent the complex motion space. However, reducing the codebook embedding dimension \(d_c\) from 768 to 128 made the VQ-VAE unable to effectively model the motion space. 

\textbf{Positional embeddings.} We notice that using absolute sinusoidal positional embeddings showed better qualitative results to relative positional embeddings when evaluating on sequence lengths seen during training, while causing motion freezing when generating sequences longer than 30 secs. This can be seen by larger \textit{FID} values in Table \ref{table:abl_studies} indicating jitter. 


\textbf{Effect of audio representations.} In our observations, we found that using Librosa\cite{librosa} features yielded similar performance to Encodec\cite{encodec} features (Table \ref{table:abl_studies}) when applied to music from the AIST++\cite{aichoreo} dataset. We also notice that motion generated using Librosa\cite{librosa} is majorly comprised of re-occurring movements for similar music segments, while Encodec features are closer to the groundtruth and is more diverse (Figure \ref{fig:comparision}). Encodec features perform significantly better on in-the-wild music. This finding is reminiscent of observations made in image-to-text generation models\cite{parti,Imagen,stablediff}, where scaling the text encoding led to a significant improvement in out-of-domain image generation quality. 




\textbf{Style Modulation.} As seen in Figure \ref{fig:comparision}, inclusion of the additional style embedding "Ballet Jazz" while conditioned on "Krump" dance style music (which has primarily upper body movements) introduces ballet (long strides) steps. However, our model is unable to take advantage of CLIPs\cite{clip} rich semantic embeddings and did not generalize to phrases beyond the training data (10 textual genres). To enhance the model's generalizability, future work could involve training the MotionSeq decoder on the HumanML3D\cite{humanml3d} dataset, which offers 45,000 text prompts.

\paragraph{Limitations.} Although we have showcased that a simple architecture can still generate high quality motions, we qualitatively observe that the generated motions do not explicitly follow choreographic rules and rythmic patterns, e.g., repeated music should have repeated dance movements. Further, current evaluation metrics are not suitable for evaluating motion quality. Text-to-motion and image generation domains use deep neural networks to extract semantically rich feature extractors while AIST++\cite{aichoreo} uses hand crafted feature extractors that are not capable of representing complex motion sequences. Further, the beat align metric is flawed as beats of the music are only a loose guide for timing and instead should use well defined choreographic rules to evaluate music-dance correspondence. 


\section{Conclusion and future work}

In this research, we present a conditional motion generation framework using a VQVAE and an autoregressive cross-conditioned transformer to generate diverse motion sequences that strongly correlate with music. We achieve state-of-the-art results on widely-used benchmarks by training the framework on a combined dataset of music-to-motion and text-to-motion. Our results showed that relative positional embeddings outperform traditional absolute embeddings on longer sequences, and Encodec\cite{encodec} music embeddings yield superior performance on in-the-wild music. Future work should focus on creating benchmarks that take advantage of already well defined choreographic rules and develop semantic feature extractors that learn mappings between the motion and music space to effectively score motion quality. By building upon this approach, future models can explore more complex architectures by improving the style modulation capabilities or introducing choreographic rules to further advance the field of music-to-motion generation.





\clearpage

\printbibliography
\end{document}